\documentclass[reprint,aps,amsmath,amssymb,amsfonts]{revtex4-1}
\usepackage{graphicx}
\usepackage{dcolumn}
\usepackage{bm}
\usepackage{txfonts}
\usepackage{multirow}
\usepackage{color}

\begin{document}

\title{Correlation in transport coefficients of hole-doped CuRhO$_2$ single crystals}

\author{K.~Kurita}
\author{H.~Sakabayashi}
\author{R.~Okazaki}

\affiliation{Department of Physics, Faculty of Science and Technology, Tokyo University of Science, Noda 278-8510, Japan}

\begin{abstract}
To clarify the origin of the nontrivial thermoelectric properties observed in the delafossite oxide CuRhO$_2$ polycrystals,
we have performed the systematic transport measurements on the single-crystalline CuRhO$_2$ samples.
In the parent compound, we find a pronounced peak structure due to a phonon-drag effect 
in the temperature dependence of the Seebeck coefficient,
which is also confirmed by the size effect experiments. 
In the Mg-substituted crystals, in contrast to the results of the polycrystals, 
both the resistivity and the Seebeck coefficient decrease with increasing Mg content $y$.
In particular, 
the coefficient $A$ for the $T^2$ term of the resistivity and 
the $T$-linear coefficient for the Seebeck coefficient at low temperatures
are well described within a simple relationship expected for metals,
which is also applicable to the correlated materials with low carrier densities.

\end{abstract}

\maketitle

\section{Introduction}

The search for the guiding principles 
toward the efficient thermoelectric materials is 
of great importance both in fundamental and applied physics.
The oxides have attracted a keen attention as 
a key class of thermoelectric materials
with
a certain merit of their high stability at high temperature in air \cite{Maignan2002,Koumoto2006,Snyder2008,Terasaki2016}.
In particular,
since the discovery of the good thermoelectric properties in 
Na$_x$CoO$_2$ \cite{Terasaki1997}, 
various cobalt oxides are investigated,
and
interesting mechanisms to enhance the 
Seebeck coefficient, $S$,
including the spin-orbital entropy of correlated $d$ electrons \cite{Koshibae2000,Koshibae2001}
and 
the peculiar shape of the electronic band structure \cite{Kuroki2007},
are proposed and examined.

The rhodium oxides are the isovalent $4d$ counterpart for the cobalt oxides,
essential to the thorough understanding of the origin of 
the thermoelectric transport properties in this system \cite{Okada2005,Klein2006,Okamoto2006,Kobayashi2007,Kobayashi2009,Shibasaki2010,Okazaki2011}.
In this respect, the delafossite CuRhO$_2$ 
is a suitable material \cite{Kuriyama2006,Shibasaki2006,Guilmeau2009,Maignan2009,Le2009,Vilmercati2013,Daou2017}, 
since the two-dimensional (2D) RhO$_2$ layer is 
consisting of the edge-shared RhO$_6$ octahedra 
to form a triangular lattice as is similar to 
the 2D triangular lattice of Co ions in Na$_x$CoO$_2$.
The parent compound is insulating and becomes conducting with the hole doping by the Mg$^{2+}$ substitution
to the Rh sites.
Indeed, polycrystalline CuRh$_{0.9}$Mg$_{0.1}$O$_2$ exhibits a large Seebeck coefficient 
with a metallic resistivity \cite{Guilmeau2009,Maignan2009},
which may originate from a ``pudding mold'' type band structure suggested by Usui \textit{et al.} \cite{Usui2009},
as is similar to the case of Na$_x$CoO$_2$.
Although there exists a contribution of the Cu layer \cite{Le2009},
combined spectroscopic experiments have shown that the Rh $4d$ electron is dominant for the conduction phenomena \cite{Vilmercati2013},
as is similar to other hole-doped delafossite compounds \cite{Yokobori2013}.

The low-temperature transport nature of CuRh$_{1-y}$Mg$_{y}$O$_2$
is,
on the other hand,
still controversial:
while the resistivity certainly varies from insulating to conducting behavior with the hole doping, 
the temperature variations of the Seebeck coefficient for $0\leq y \leq 0.2$
fall into a single curve below 150~K \cite{Shibasaki2006},
highly incompatible with the conventional behavior where 
the Seebeck coefficient decreases in magnitude with increasing carrier density \cite{Snyder2008}.
Since the Seebeck coefficient measures 
the temperature derivative of the chemical potential,  
$\partial \mu/\partial T$ \cite{Peterson2010,Yamamoto2017},
the observed $y$-independent Seebeck coefficient indicates that
the compressibility, $\partial n/\partial \mu$ ($n$ being the carrier density),
is divergently large \cite{Shibasaki2006}.
This implies an occurrence of a phase separation,
in which a percolation network for the doped holes is away from the substituted Mg ions.
Thus it represents an intriguing 
doping effect to be further addressed where the substituted ions act as the carrier donors/acceptors with weak scattering,
although the earlier studies are performed by using polycrystalline samples.

The aim of this paper is to clarify the low-temperature thermoelectric properties of CuRh$_{1-y}$Mg$_{y}$O$_2$
by using single-crystalline samples.
We find that the parent compound shows a peak structure in the temperature dependence of the 
Seebeck coefficient at around 120~K.
We also observe a size effect of the peak structure, indicating an 
existence of the phonon-drag effect. 
In the Mg-substituted crystals, the Seebeck coefficient systematically varies with the hole doping unlike the polycrystalline data.
Moreover, 
we show that 
the relation between 
the $T$-linear coefficient for the Seebeck coefficient, $S/T$, and the coefficient $A$ for the $T^2$ term in the Fermi-liquid resistivity 
is described within a simple model for metals.
In contrast to other universal relations, such as the Kadowaki-Woods ratio $A/\gamma^2$ ($\gamma$ being the electronic specific heat coefficient) \cite{Kadowaki1986}
and $S/\gamma T$ ratio suggested by Behnia \textit{et al.} \cite{Behnia2004}, 
the present relation among the transport coefficients is widely applicable to the correlated materials 
even with low carrier density.

\section{Experiments}

The experiments were performed using 
CuRh$_{1-y}$Mg$_y$O$_2$ single crystals ($y=0,0.03,0.06,0.1$). 
Powders of Rh$_2$O$_3$ (99.9\%), MgO (99.99\%), and CuO (99.9\%)
were mixed in a stoichiometric ratio and
calcined in air at 1203~K for 36~h.
The powders were then ground and calcined again in air at 1203~K for 72~h \cite{Shibasaki2006}.
The single crystals were then grown by a self-flux method as is similar to the crystal growth of CuAlO$_2$ \cite{Koumoto2001}.
CuO (99.9\%) powder was added 
with the calcined powder as a flux. 
The concentration of CuRh$_{1-y}$Mg$_y$O$_2$ powder was set to be 10\% in molar ratio. 
The mixture was put in a platinum crucible and 
heated up to 1423~K in air with a heating rate of 200~K/h.
After keeping 1423~K  for 10~h, 
it was slowly cooled down  with a rate of 0.5~K/h, 
and at 1323~K, the power of the furnace was switched off.
As-grown samples were rinsed in 1~M HNO$_3$ to remove the flux,
and then annealed at 673~K for 20~h in air 
to obtain the homogeneous samples.
Typical crystal dimension was $1\times0.3\times 0.01$~mm$^3$.

\begin{figure}[t]
\begin{center}
\includegraphics[width=1\linewidth]{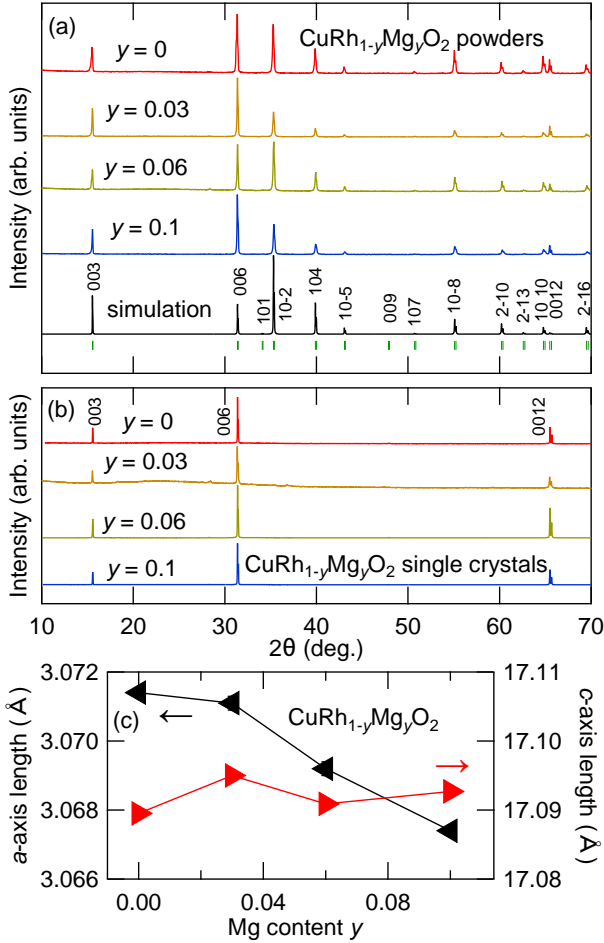}
\caption{
X-ray diffraction patterns of 
(a) CuRh$_{1-y}$Mg$_y$O$_2$ powders and
(b) single-crystalline samples.
The powders are obtained by grinding the single crystals.
The vertical bars show the Bragg positions.
(c) The $a$-axis (left) and the $c$-axis (right) length as 
a function of the Mg content $y$.
}
\end{center}
\end{figure}

X-ray diffraction (XRD) measurements were performed at room temperature
by using an x-ray diffractometer (Rigaku UltimaIV) with Cu K$\alpha$ radiation in a $\theta$-2$\theta$ scan mode. 
For the powder XRD experiments, we ground the single crystals grown by the self-flux method.
In the case of the single-crystalline experiments, the scattering vector was normal to the surface of the sample. 
The in-plane resistivity was measured with a conventional dc four-probe method.
The excitation current of $I=20~\mu$A was provided by a Keithley 6221 current source and the sample voltage was measured with a synchronized Keithley 2182 nanovoltmeter. 
These two instruments were operated in a built-in Delta mode to cancel unwanted thermoelectric voltage.
The Seebeck coefficient was measured by using a steady-state method with a typical temperature gradient of 0.5~K/mm made by a resistive heater. 
The thermoelectric voltage of the sample was measured with Keithley 2182A nanovoltmeter.
The temperature gradient was measured with a differential thermocouple made of copper and constantan in a liquid He cryostat. 
The thermoelectric voltage from the wire leads was subtracted.

\section{Results and discussion}

\subsection{Structural and transport properties}

Figures 1(a) and 1(b) represent the x-ray diffraction patterns of 
CuRh$_{1-y}$Mg$_y$O$_2$ powders and single-crystalline samples, respectively.
No impurity phase is detected in the powder data, 
and only $(00l)$ reflections are observed in the 
single-crystalline samples, 
showing that the measured crystal surface is the $ab$ plane.
Figure 1(c) depicts the Mg content $y$ dependence of the lattice parameters.
The $a$-axis length decreases with increasing $y$, 
while the $c$-axis length is almost constant for $y$,
as is consistent with the previous reports \cite{Shibasaki2006}.
The volume reduction with increasing $y$ is explained by the comparison of the 
ionic radii for Mg$^{2+}$ ($0.72$ $\AA$), Rh$^{3+}$ ($0.665$ $\AA$), and
Rh$^{4+}$ ($0.60$ $\AA$) \cite{Shannon1976}:
If we simply express the valence state of CuRh$_{1-y}$Mg$_{y}$O$_2$ as 
CuRh$^{3+}_{1-2y}$Rh$^{4+}_{y}$Mg$_{y}$O$_2$ from the charge neutrality,
the average ionic radius of the Rh site $R_{\rm avg}$ is calculated as
$R_{\rm avg} = 0.665\times(1-2y) + 0.66\times2y$. 
The reduction of $R_{\rm avg}$ at $y = 0.1$ is then calculated as 0.15\%. 
On the other hand, 
experimental reduction of the $a$-axis length at $y = 0.1$ is 0.13\%, 
which well agrees with the theoretical calculation.

\begin{figure}[t]
\begin{center}
\includegraphics[width=1\linewidth]{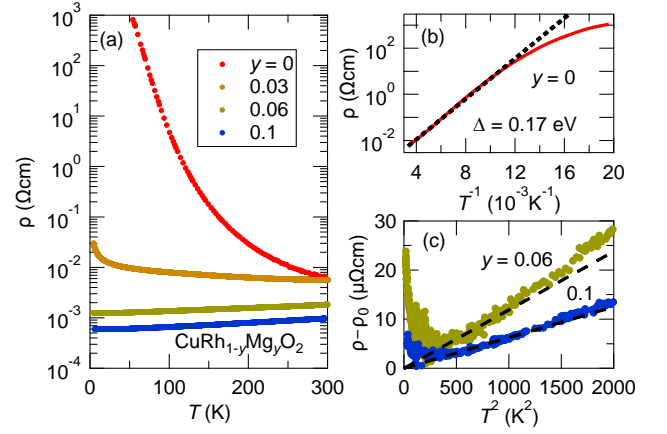}
\caption{
(a) Temperature dependence of the in-plane resistivity, $\rho(T)$,
of CuRh$_{1-y}$Mg$_y$O$_2$ crystals.
(b) $T^{-1}$ dependence of $\ln\rho$ for CuRhO$_2$.
The dotted line represents a thermal activation formula.
(c) $T^2$ dependence of $\rho-\rho_0$ for the doped compounds.
The dashed lines represent the fitting results.
}
\end{center}
\end{figure}

Figure 2(a) shows the 
temperature variations of the in-plane resistivity, $\rho(T)$,
for CuRh$_{1-y}$Mg$_y$O$_2$ crystals ($0\leq y\leq 0.1$).
The resistivity systematically decreases with the Mg substitution,
indicating the successful hole doping.
Figure 2(b) depicts $\ln\rho$ as a function of $T^{-1}$ for the parent compound,
and the dotted line represents a thermal activation fitting 
$\rho\propto\exp(\Delta/2k_BT)$,
where $\Delta = 0.17$~eV is the activation energy.
Now this value is smaller than the energy gap value of 0.8~eV
estimated from the band calculations \cite{Usui2009,Maignan2009},
indicating a thermal excitation from the valence band to
an acceptor level.
Below 100 K, the resistivity deviates from the activation formula 
since an impurity band conduction sets in.
In the metallic samples for $y \geq 0.06$,
as shown in Fig.~2(c),
the low-temperature resistivity shows a 
Fermi-liquid behavior $\rho(T) = \rho_0 +AT^2$ with
a small upturn below 20~K,
as is similar to the polycrystalline study \cite{Maignan2009} and to
the results of other layered oxides \cite{Masset2000,Limelette2005,Hsieh2014,Ikeda2016}.
The doping dependence of $A$ will be mentioned later.

\begin{figure}[t]
\begin{center}
\includegraphics[width=1\linewidth]{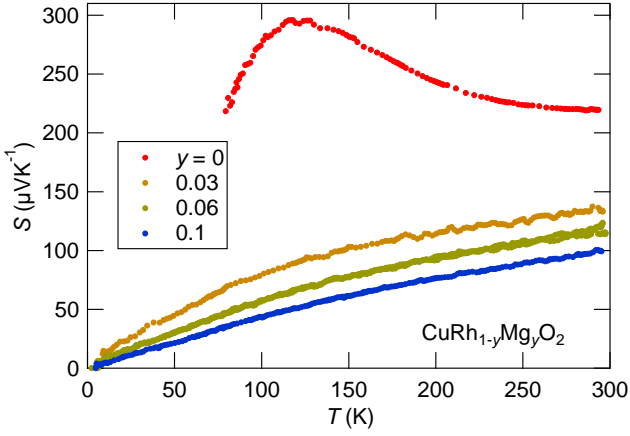}
\caption{
Temperature dependence of the in-plane Seebeck coefficient, $S(T)$,
of CuRh$_{1-y}$Mg$_y$O$_2$ crystals.
}
\end{center}
\end{figure}

The temperature variations of the Seebeck coefficient are displayed in
Fig. 3.
In sharp contrast to the polycrystalline results \cite{Shibasaki2006}, 
the Seebeck coefficient systematically varies with the Mg content $y$.
In the parent compound, it exhibits a relatively large value of $S\sim200$~$\mu$V/K 
at room temperature and shows a peak structure at $T\simeq120$~K.
Such a structure may originate from a phonon-drag effect, 
because it is not resolved in the polycrystalline sample with
a considerable phonon scattering effect at the grain boundaries \cite{Masui2002}.
Also note that the observation of the phonon-drag effect supports 
the thermal activation transport in the resistivity as mention above,
rather than the variable range hopping conduction.
For $y\geq 0.03$, the Seebeck coefficient shows a metallic behavior, and 
the overall temperature dependence of $S$ qualitatively agrees with the 
theoretical calculations \cite{Usui2009}.
The low-temperature $T$-linear coefficient decreases with the doping, 
which will be argued with the results of the resistivity.

\subsection{Phonon-drag effect}

\begin{figure}[t]
\begin{center}
\includegraphics[width=1\linewidth]{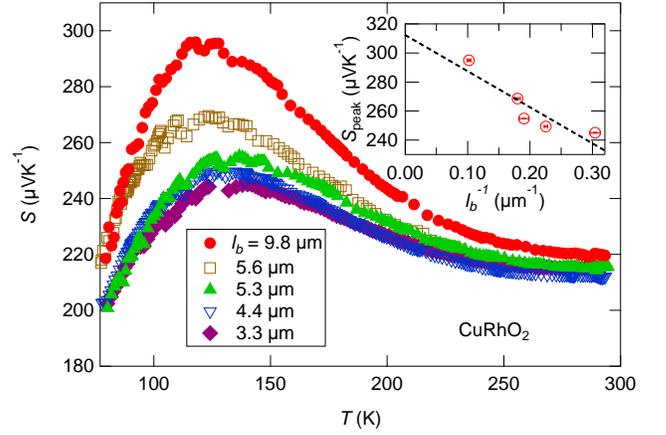}
\caption{
Temperature dependence of the in-plane Seebeck coefficient, $S(T)$,
of several CuRhO$_2$ crystals with different dimensions.
The characteristic length $l_b$ is 
a mean free path dominated by the crystal boundary scattering,
which is 
obtained from the sample dimension 
as is described in the main text. 
Inset: the Seebeck coefficient at the peak temperature, $S_{\rm peak}$,
as a function of $l_b^{-1}$.
The dashed line is a fitting result.
}
\end{center}
\end{figure}

To discuss the phonon-drag effect in the CuRhO$_2$ crystal in more details, 
we have examined the size effect on the Seebeck coefficient.
It is well known that the phonon-drag contribution to the Seebeck effect can be 
enhanced by sizing up the crystal owing to the enhancement of the mean free path of the ballistic phonon.
Figure 4 represents the temperature variations of the Seebeck coefficient
of several CuRhO$_2$ crystals with different dimensions.
Here, a mean free path dominated by the crystal boundary scattering, $l_b$, 
can be estimated by using the sample dimensions as \cite{McCurdy1970}
\begin{eqnarray}
l_b &=& 
\left(\frac{1}{4}Dx^{1/2}\right)
\left[
3x^{1/2}\ln\left\{x^{-1}+\left(x^{-2}+1\right)^{1/2}\right\}\right. \nonumber \\ 
&+&
3x^{-1/2}\ln\left\{x+\left(x^{2}+1\right)^{1/2}\right\} \nonumber \\ 
&-&\left.\left(x+x^3\right)^{1/2} + x^{3/2}
-\left(x^{-1}+x^{-3}\right)^{1/2} + x^{-3/2}\right],
\end{eqnarray}
where $D\times xD$ is the cross section of the crystal with a rectangular shape.
As seen in Fig.~4, the peak structure in $S(T)$ is enhanced with increasing $l_b$,
implying the phonon-drag effect.

It should be noted that, however, other scattering processes such as 
the phonon-phonon scattering should be dominant in the present
temperature range,
in contrast to the phonon-drag effect at low temperature \cite{Takahashi2016}.
Let $l_T$ be a temperature-dependent mean free path,
which is determined by the other scatterings,
and then the phonon mean free path, $l_{\rm ph}$, 
is generally given as
\begin{eqnarray}
l_{\rm ph}^{-1} =l_{T}^{-1} + l_{b}^{-1}.
\end{eqnarray}
Here, $l_{T} \ll l_{b}$ holds around $T=120$~K, since the phonon transport should be diffusive. 
Now, the phonon-drag contribution to 
the Seebeck coefficient, $S_{\rm ph}$, is given as \cite{Herring1954}
\begin{eqnarray}
S_{\rm ph} =\frac{\beta v l_{\rm ph}}{\mu_e T},
\end{eqnarray}
where $\beta$, $v$, and $\mu_e$ are an electron-phonon coupling constant ($0<\beta<1$), 
the phonon velocity, and the carrier mobility, respectively.
Then the total Seebeck coefficient is approximately expressed as
\begin{eqnarray}
S \simeq S_d + \frac{\beta v l_{T}}{\mu_e T}\left(1-\frac{l_{T}}{l_{b}}\right),
\label{seebeckph}
\end{eqnarray}
where $S_d$ is the diffusive term of the Seebeck coefficient.
The inset of Fig.~4 depicts the maximum Seebeck coefficient at the 
peak structure, $S_{\rm peak}$, as a function of $1/l_b$.
According to Eq. \ref{seebeckph}, 
the experimental data are roughly fitted by a linear function of $1/l_b$ as shown by the 
dashed line, supporting the present phonon-drag model.

\subsection{Relation in the transport coefficients}

\begin{figure}[t]
\begin{center}
\includegraphics[width=1\linewidth]{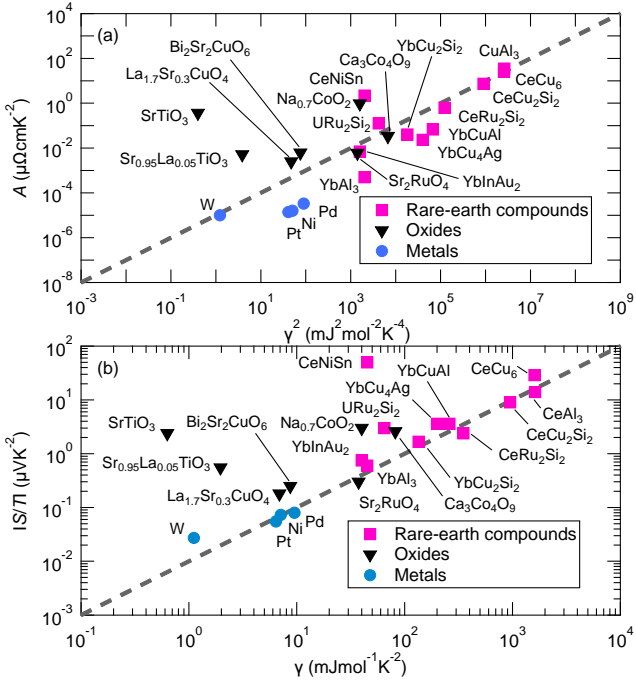}
\caption{
(a) $A$ vs. $\gamma^2$ and
(b) $|S/T|$ vs. $\gamma$
plots in log-log scale for various materials \cite{Jaccard1985,Aliev1984,Bredl1984,Sparn1985,Sato1987,Loneysen1994,Hiess1994,Izawa1999,Terashima2002,Amato1988,Lacerda1989,Sakurai1988,Maple1986,Jaccard1982,Havinga1973,
AlamiYadri1999,Sales1976,Casanova1990,Bauer1991,Jaccard1985Yb,Pott1981,Besnus1986,
Konstantinovic2002,Mayama1998,Saito2017,Elizarova2000,Nakamae2003,Lee2006,Yoshino1996,Maeno1997,Okuda2001,Farrell1970,White1959,Fletcher1970,Mizutani1977,Fletcher1968,Trodahl1973,Andres1975,Nakamoto1995,Kambe1996,Palstra1986,Ebihara2000,Thompson1987,Tsujii1997,Mirzaei2013,Li2004,White1959,Rice1968,Budworth1960,Waite1956}.
The dashed lines are guides to the eye, showing that $A/\gamma^2$ and $|S/T|/\gamma$ are 
constants.
The deviations are clearly seen in the low-carrier materials.
}
\end{center}
\end{figure}

Now let us discuss the doping dependence of the resistivity and the Seebeck coefficient.
As shown in Fig.~2(c) and Fig.~3, 
both the coefficient $A$ for the $T^2$ term of the resistivity and 
the $T$-linear coefficient for the Seebeck coefficient at low temperatures
are reduced with increasing Mg content $y$,
which is readily understood with the increase in the carrier density $n$ \cite{AM}.

Here, let us compare the present results with other systems.
It is well known that both $A$ and  $S/T$ values  are correlated with the 
electronic specific heat coefficient, $\gamma$, 
as is represented by
Kadowaki-Woods \cite{Kadowaki1986} and Behnia plots \cite{Behnia2004}.
Figures~5(a) and 5(b) represent the 
$A$ vs. $\gamma^2$ and $|S/T|$ vs. $\gamma$
plots, respectively, for various materials \cite{Jaccard1985,Aliev1984,Bredl1984,Sparn1985,Sato1987,Loneysen1994,Hiess1994,Izawa1999,Terashima2002,Amato1988,Lacerda1989,Sakurai1988,Maple1986,Jaccard1982,Havinga1973,
AlamiYadri1999,Sales1976,Casanova1990,Bauer1991,Jaccard1985Yb,Pott1981,Besnus1986,
Konstantinovic2002,Mayama1998,Saito2017,Elizarova2000,Nakamae2003,Lee2006,Yoshino1996,Maeno1997,Okuda2001,Farrell1970,White1959,Fletcher1970,Mizutani1977,Fletcher1968,Trodahl1973,Andres1975,Nakamoto1995,Kambe1996,Palstra1986,Ebihara2000,Thompson1987,Tsujii1997,Mirzaei2013,Li2004,White1959,Rice1968,Budworth1960,Waite1956}.
Note that we refer the literature data, but the analysis for the Fermi-liquid behavior is a remaining issue \cite{Tateiwa2012,Jaoui2018}.
Here, although the Kadowaki-Woods plot can be corrected by involving 
the degeneracies \cite{Tsujii2003,Kontani2004,Tsujii2005}, 
both plots well explain the universal relations among those quantities for the correlated materials
with a metallic carrier density.
On the other hand, 
since both $A/\gamma^2$ and $S/\gamma T$ ratios
include the carrier density $n$ \cite{Behnia2004,Takimoto1996,Continentino2000},
the data of correlated materials with low carrier density, 
such as
a Kondo insulator CeNiSn \cite{Terashima2002} 
and slightly doped SrTiO$_3$ \cite{Okuda2001},
significantly deviate from the universal relation 
in these plots.
Note that
there are several attempts to correct this deviation due to the carrier density \cite{Hussey2005,Jacko2009}.

\begin{figure}[t]
\begin{center}
\includegraphics[width=1\linewidth]{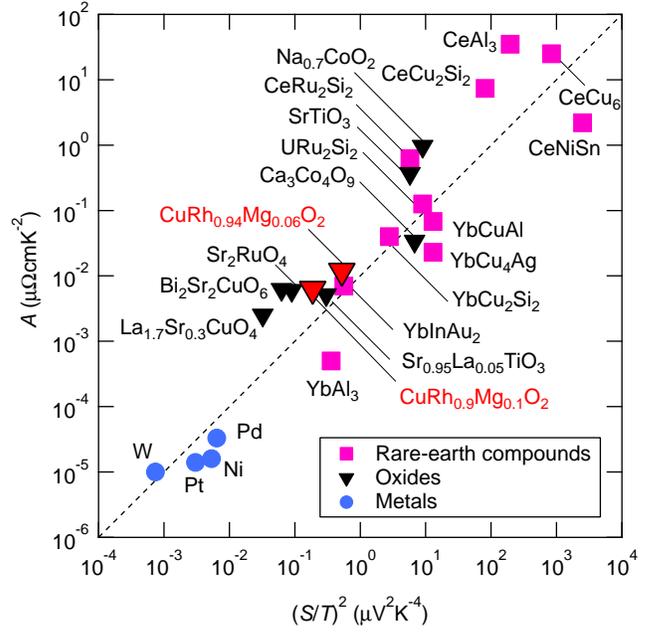}
\caption{
$A$ vs. $(S/T)^2$ plot for various materials.
The dashed line shows a proportional relation among the transport coefficients, $A$ and $(S/T)^2$, 
with $l_{\rm quad}=4$~nm in Eq. \ref{ratioAS}.
}
\end{center}
\end{figure}

In Fig.~6, we instead plot $A$ as a function of $(S/T)^2$ for the materials shown in Figs.~5(a) and 5(b),
in addition to the present data for CuRh$_{1-y}$Mg$_{y}$O$_2$ ($y=0.06, 0.1$).
All the data, including the low-density materials mentioned above, seem to fall into a single line.
Now, 
the $A$ term in the resistivity is expressed as \cite{MottMI,Lin2015}
\begin{eqnarray}
A =\frac{\hbar}{e^2}\left(\frac{k_B}{\varepsilon_F}\right)^2l_{\rm quad},
\end{eqnarray}
where $\hbar$ is the reduced Planck constant,
$e$ the elementary charge, and 
$\varepsilon_F$ the Fermi energy.
Here $l_{\rm quad}$ is a characteristic length for the scattering events, 
which is given as $l_{\rm quad} = k_F\sigma_{\rm cs}$
($k_F$ and $\sigma_{\rm cs}$ are the Fermi wavenumber and
the collision cross section between two electrons, respectively).
On the other hand, in a free-electron model,
the Seebeck coefficient divided by temperature is given as
\begin{eqnarray}
\frac{S}{T} = - \frac{\pi^2}{2}\frac{k_B}{e}\frac{k_B}{\varepsilon_F}.
\end{eqnarray}
Therefore, 
we express the ratio as
\begin{eqnarray}
\frac{A}{(S/T)^2} = \frac{4}{\pi^4}\frac{\hbar}{k_B^2}l_{\rm quad},
\label{ratioAS}
\end{eqnarray}
in which the Fermi energy is cancelled out.
The present ratio $A/(S/T)^2$ is, in other words, a direct measure for $l_{\rm quad}$.
This ratio can also be expressed as $(2/3\pi^3)(l_{\rm quad}/G_{\rm en})$,
using the thermal conductance quantum, $G_{\rm th}$, divided by temperature,
$G_{\rm th}/T = G_{\rm en} = (\pi/6)(k_B^2/\hbar)$.
Since $G_{\rm en}$ ($=G_{\rm th}/T$) may be related to an entropy conductance, 
$A/(S/T)^2\simeq l_{\rm quad}/G_{\rm en}$ may imply a resistivity for an entropy flow, 
in analogy to the Fermi-liquid resistivity $AT^2 \simeq (l_{\rm quad}/G_0)(k_BT/\varepsilon_F)^2$ 
($G_0$ being the conductance quantum), which expresses a resistivity for the charge flow. 
Here, in Eq. \ref{ratioAS},
$l_{\rm quad}$ is indeed material-dependent but 
mostly lies in a narrow range between 1 and 40~nm \cite{Lin2015},
consistent with the plot shown in Fig.~6,
in which various experimental data are on a single line
calculated with a constant $l_{\rm quad}$ of 4~nm.
It is an interesting issue to clarify the origin of the weak material dependence of $l_{\rm quad}$
for various compounds with very different size and shape of Fermi surfaces.

Noted that, in the present compound CuRh$_{1-y}$Mg$_{y}$O$_2$,
a pudding-mold type band structure is suggested to enhance the Seebeck coefficient \cite{Usui2009}.
The value of $S/T$ in Fig.~6 is, however, evaluated at low temperature,
at which a simple metallic model can be applicable.
Also, it is unlikely to apply the high-temperature spin-orbital entropy model \cite{Nishina2017,Takahashi2018}.
For the present ratio $A/(S/T)^2$, an interesting exception may be 
a material around the Lifshitz transition, at which the Seebeck coefficient becomes divergently large 
while the resistivity remains intact \cite{Varlamov1989,Ito2019}.

\section{Summary}

In this study, we have carried out the transport measurements on the delafossite oxides
CuRh$_{1-y}$Mg$_{y}$O$_2$ single crystals, 
and find the systematic doping variations of the resistivity and the Seebeck coefficient.
Particularly, 
we find a correlation between 
the coefficient $A$ for the $T^2$ term of the Fermi-liquid resistivity and 
the $T$-linear coefficient for the Seebeck coefficient at low temperature,
which can be described within a simple relationship expected for metals.
This relation may also be applicable to the correlated materials with low carrier densities.

\begin{acknowledgments}
This work was supported by JSPS KAKENHI Grants No. JP17H06136, No. JP18K03503, and No. JP18K13504.

\end{acknowledgments}

\end{document}